\documentstyle[12pt,epsfig,wrapfig]{article}
\include{times,mathptm}
\renewcommand{\section}[1]{\vspace{6pt} \noindent\mbox{#1} \newline \noindent}
\renewcommand{\subsection}[1]{\vspace{6pt} \noindent\mbox{\underline{#1}} 
\newline \noindent}
\renewcommand{\subsubsection}[1]{\vspace{6pt} \noindent\mbox{\underline{#1}}
\noindent}

\newfont{\sansb}{cmssbx10}
\newfont{\sans}{cmss10}

\begin{document}
{\small HE 4.1.11 \vspace{-24pt}\\}     
{\center \LARGE A SEARCH FOR VERY HIGH ENERGY NEUTRINOS FROM ACTIVE 
GALACTIC NUCLEI
\vspace{6pt}\\}
J.~W.~Bolesta$^1$, H.~Bradner$^9$, U.~Camerini$^{12}$, J.~Clem$^7$, S.~T.~Dye$^8$,
J.~George$^{11}$, P.~W.~Gorham$^{13}$, P.~K.~F.~Grieder$^6$, J.~Hauptman$^5$, T.~Hayashino$^3$,
M.~Jaworski$^{12}$, T.~Kitamura$^4$, S.~Kondo$^1$, J.~G.~Learned$^1$, R.~March$^{12}$,
T. ~Matsumoto$^3$, S.~Matsuno$^1$, K.~Mauritz$^5$, P.~Minkowski$^6$, T.~Narita$^{12}$,
D.~J.~O'Connor$^1$, Y.~Ohashi$^2$, A.~Okada$^2$, V.~Peterson$^1$, V.~J.~Stenger$^1$, S.~Uehara$^4$,
M.~Webster$^{10}$, R.~J.~Wilkes$^{11}$, K.~K.~Young$^{11}$, A.~Yamaguchi$^3$ \vspace{6pt}\\
{\it $^1$University of Hawaii, USA}
{\it $^2$Institute of Cosmic Ray Research, University of Tokyo, Japan}
{\it $^3$Tohoku University, Japan}
{\it $^4$KEK, Japan}
{\it $^5$Iowa State University, USA}
{\it $^6$University of Bern, Switzerland}
{\it $^7$Bartol Research Institute, USA}
{\it $^8$Hawaii Pacific University, USA}
{\it $^9$Scripps Institute of Oceanography, USA}
{\it $^{10}$Vanderbilt University}
{\it $^{11}$University of Washington, USA}
{\it $^{12}$University of Wisconsin, USA}
{\it $^{13}$NASA Jet Propulsion Laboratory, USA} \vspace{-12pt}\\
{\center ABSTRACT\\}
We report the results of a search for neutrino-induced particle cascades
using a deep ocean water ${\rm \check{C}erenkov}$ detector. 
The effective mass of the detector, a string
of seven 40 cm diameter photomultipliers (PMT) at 5.2 m spacing, is found through
simulation analysis to be surprisingly large: 
greater than $10^6$ tons of water at incident neutrino energies of $10^6 ~GeV$.
We find no evidence for neutrino-induced cascades in 18.6 hours of observation. 
Although the limit implied by this observation is the strongest yet for
predictions of active galatic nuclei (AGN) neutrinos at energies above 100 TeV,
perhaps the more intriguing result is that the power of these techniques can be exploited 
to test these AGN models in a relatively short time.

\setlength{\parindent}{1cm}
\section{INTRODUCTION AND MOTIVATION}
The idea behind our measurement of the neutrino flux is in essence,
the same as extensive air shower (EAS) detectors, except in the ocean.
Due to the optical properties of the deep ocean and the fact that
water is $\sim 10^3$ more dense than the atmosphere, detector livetimes
can accrue at a rate of $\sim$ kiloton-years (kty) {\it per day.}

We have re-analyzed data taken in November 1987
with respect to the optical properties of the
deep ocean and used the results in our monte carlo. We then searched the
data for the presence of cascades. In the following we present only the results
of our analysis, the details of which can be found in Bolesta et al. (1997). 

\section{EFFECTIVE MASS FOR CASCADE DETECTION}
\subsection{Background Reduction}
According to the models of Bierman (1992), Szabo and Protheroe (1994),
Stecker and Salomon (1995), Protheroe (1996), we determined an
{\it a priori} selection criterion for AGN neutrino-induced cascades
based on our simulations:
the event must produce $\geq 5$ photoelectrons (PE) per PMT, for 6 out of 7 PMTs.
After detailed
simulations of detector response and all backgrounds, this cut was found to
reject $\geq 90\%$ of the atmospheric muons, 
while still accepting $\sim 70\%$ of all
of the cascades. Thus we maintained this selection criterion throughout
the analysis.

\subsection{Effective Mass Determination}
The effective mass is calculated by taking the ratio of detected to generated
monte carlo events as a function of the radial distance from the detector,
and integrating it out to the edge of the generation volume. 
This estimate is shown in Figure 1. The effective mass 
exceeds 10 Mtons at the energy of the W resonance (Glashow, 1960),
and is of order 2 Mtons at 100 TeV.
At the highest energies, it approaches 200 Mtons, or
20\% of a cubic kilometer of water.

\begin{wrapfigure}{r}{8cm}
\epsfig{file=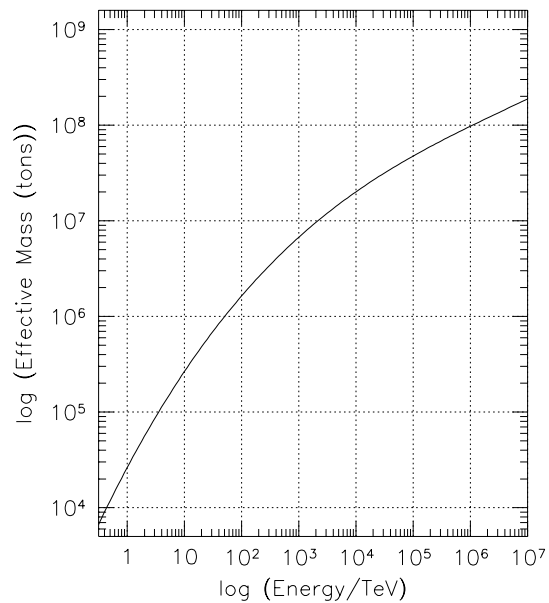,width=8.5cm,height=8.5cm,angle=0}
\caption{ Effective mass of the SPS for cascade detection as a 
function of cascade energy.}
\end{wrapfigure}

\section{RESULTS}
\subsection{Data Selection}
To establish a pure sample of well-constrained muon events to act as a
standard against which we can compare possible cascade candidates, we
cut events which had less than 6 PMTs above
the 1.6 PE threshold to reduce noise contamination.
The remaining events constituted
our parent sample of what is expected to be mostly muon events with a
possible sub-population of cascade events.

\subsection{Muon fitting}
The muon track parameters were estimated from the data using
maximum likelihood.The cumulative probability curve for this likelihood function
was estimated
by a Monte Carlo integration, and fits which fell outside of $\sim 2\sigma$
($\sim 4\%$ probability) were excluded to avoid contamination from events
that contained pre-pulse,
bremsstrahlung, or bioluminescence activity.
The fitting efficiency after this cut was $\sim 70\%$.

\begin{wrapfigure}[18]{r}{8cm}
\begin{picture}(260,180)(0,8)
\epsfig{file=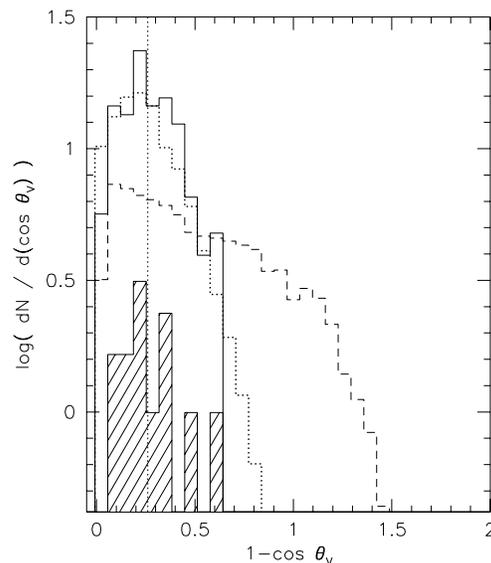,width=8.0cm,height=8.0cm,angle=0}
\end{picture}
\caption{ Angular distribution of events after fitting to a
cascade hypothesis. The vertical
dotted line is the ${\check{C}erenkov}$ angle for muons.}
\end{wrapfigure}

For our observation, the Monte Carlo estimate predicts that 8.8\% 
of atmospheric muons which pass the 1.6 PE cut should also pass
the 5 PE cut, corresponding to 18.7 events.
We observed 17, consistent with no AGN neutrino cascade events present.
All of the 17 events
can be fit to parameters consistent with atmospheric muons.

\subsection{Testing for possible cascades}
To provide an independent test for the presence of cascades in the
high--PE sample,
we used a cascade likelihood fitting
routine similar to that used to fit the muon track parameters applied
both to the entire probable muon event parent sample, and to the subsample
of those events above the 5 PE threshold. 

The results of fitting both the parent muon sample and the high PE 
subsample
to the cascade hypothesis are shown in Figure 2, where the cascade vertex
angular distribution
of  $(1 - cos~{\theta}_V)$ is plotted for both samples. 
The plot shows that the cascade fits of the parent muon sample (solid line) favor
a range of zenith angles clustered around the muon ${\rm \check{C}erenkov}$
angle for vertical tracks. 
The high PE distribution (hatched) shows no deviation from the
parent distribution. Monte Carlo analysis of the fitting errors
shows that the standard deviation for these fits in $\cos \theta$ is
$0.17$. Thus both samples appear to be dominated by nearly vertical muon
events, consistent with expectations. For comparison, we have also fitted
a set of simulated atmospheric muon events (dotted histogram) and these
show a distribution consistent with the data.
Also plotted in Figure 2 is the fitted distribution of vertex
angles for simulated cascades from a typical AGN model,
normalized to the number of events in the $\geq 1.6$ PE sample.
We have used
the matter attenuation models of Gandhi et al. (1996)
to determine the effect of earth attenuation. It is clear that the
fitted events do not appear to be drawn from the cascade distribution,
either in the high PE sample or the larger parent sample.

No evidence for any AGN--neutrino--induced cascades is found
in our data, either from the presence of an excess above background,
or from the angular distribution of the events which comprise the most
likely candidates for cascades. In all cases the data are completely consistent
with atmospheric cosmic--ray muon events.

\begin{wrapfigure}{r}{8cm}
\epsfig{file=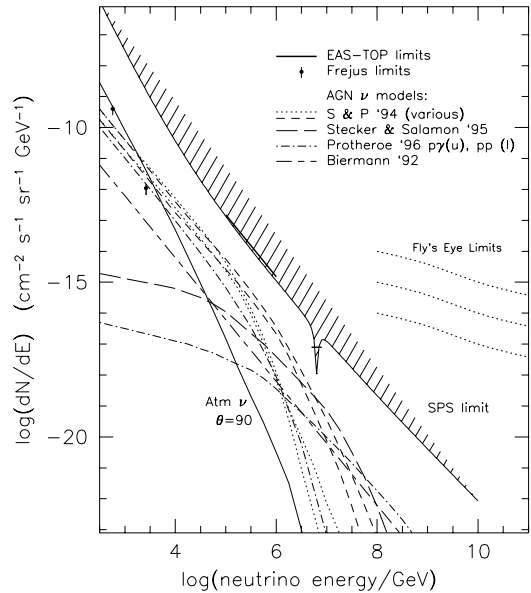,width=8.5cm,height=8.5cm,angle=0}
\caption{The limit derived here (``SPS limit'') is plotted along with
a number of neutrino models and other limits. See text for details.}
\end{wrapfigure}

\subsection{Limits on AGN neutrinos}
We establish the limit, plotted in Figure 3, by the 
90\% confidence level Poisson--statistics prescription of
a maximum allowed signal flux of 2.3 counts above our estimated minimum
detectable flux. This provides a limit that may be compared with the EAS-TOP
limit (Aglietta et al. 1994), in which 7 events were observed with 11
expected, also consistent with background.

If we use a more conservative approach to establishing the limit, by
assuming that the 17 observed events could include a possible signal
(though no evidence was seen for this in event--by--event analysis)
then the 90\%CL limits are 2.6 times higher. However we
note that this prescription is not consistent with the approach of
Aglietta et al. (1994) and the limits cannot be accurately compared
in this case.

Because the typical AGN neutrino model predicts that the ratio of the number of
electron neutrinos to muon neutrinos (and antineutrinos) is of order
$1/2$, we have estimated the differential limit under this assumption.

In Figure 3 we also plot a number of suggested models, as
well as atmospheric neutrino spectra, and limits from the 
underground muons observed by the Frejus
collaboration (Rhode et al. 1996) and the extensive air-shower
limits given by the EAS-TOP experiment  (Aglietta et al 1994; 1995).
The limits from the Fly's Eye (Baltrusaitas et al. 1985) apply for
three different assumed cross sections for $\nu_e$ interactions in the
atmosphere: $\sigma_\nu = 10^{-31}~cm^{2}$ (uppermost limit),
$10^{-30}~cm^{2}$ (middle limit),and $~10^{-29}~cm^{2}$ (lowest limit).
In each case the cross section is assumed constant for $10^{8}\leq E_\nu \leq
10^{11} ~GeV$. The cross sections used by Baltrusaitas et al.
appear to be about an order of magnitude 
higher than those estimated by Gandhi et al (1996).
 
The AGN neutrino 
models shown are from Szabo and Protheroe (1994;  S \& P in the figure),
Stecker and Salamon (1995), Protheroe (1996), Biermann (1992) and are
a representative sample of those available. The Frejus limits, based on
measurements of horizontal muon events,
appear to eliminate the highest flux models of Szabo and Protheroe (1994).
Also shown here is the expected horizontal flux from atmospheric neutrinos,
from Lipari (1993).

The limit at the W resonance includes the
typical ratio of electron to muon neutrinos in the models and
the severe attenuation
of these electron neutrinos passing through the Earth. The limit
at this energy is:
\begin{equation}
{{dF_{\nu}} \over {dE_{\nu}}} (6.3~PeV; AGN~ \nu)~ \leq 1.1 \times 10^{-18} 
cm^{-2} s^{-1} ~sr^{-1}~GeV^{-1}.
\end{equation}
This is the most stringent limit at this energy
and improves on the existing EAS-TOP
limit by about a factor of 7. 

Our value for the limit assumes that the
electron neutrino+antineutrino to muon neutrino+antineutrino ratio is 
$\sim 0.5$, as most models predict, and that there is no significant
particle--antiparticle asymmetry.
At the resonance energy, the model--{\it independent}
limit for anti-electron neutrinos is more stringent:
\begin{equation}
{{dF_{\nu}} \over {dE_{\nu}}} (6.3~PeV; mod.~indep.)~ \leq 3.2 \times 10^{-19} 
cm^{-2} s^{-1} ~sr^{-1}~GeV^{-1}.
\end{equation}

\section{CONCLUSION}
We find it quite remarkable that a modest detector can achieve such
a large sensitive mass, though not designed nor optimized for this use.
This instrument, intended merely as a proof-of-concept
for a deep-ocean muon tracking instrument, has within
less than 1 day's total livetime produced the first limits at these 
high energies which begin to approach the predictions of
the AGN neutrino models.

\section{ACKNOWLEDGEMENTS}
We wish to thank Diane Ibaraki for her
help recovering the data, and Sandip Pakvasa for useful discussions. This work 
was supported in part by the U.S. Department of Energy.

\section{REFERENCES}
\setlength{\parindent}{-5mm}
\begin{list}{}{\topsep 0pt \partopsep 0pt \itemsep 0pt \leftmargin 5mm
\parsep 0pt \itemindent -5mm}
\vspace{-15pt}

\item Aglietta,~M. {et al.},
{\it Physics Letters} {\bf B333}, 555-560, (1994).
 
\item Aglietta,~M. {et al.},
in {\it Proceedings of the 24th International Cosmic Ray Conference},
(Rome: IUPAP) Vol. 1, 638, (1995).

\item Babson,~J.~F. {et al.},
{\it Physical Review} {\bf D42}, 3613, (1990).

\item Baltrusaitas,~R.~M. {et al.},
{\it Physical Review} {\bf D31}, 2192, (1985).

\item Biermann,~P.,
in {\it Proceedings of the Workshop on High Energy Neutrino Astrophysics (HENA)},
V.~Stenger, J.~Learned, S.~Pakvasa and X.~Tata,
eds.,  Hawaii, 23-26 March 1992,
(Singapore: World Scientific), 86, (1992).

\item Bolesta,~J.~W., {et al.},
submitted to {\it The Astrophysical Journal}, (1997).

\item Gandhi,~R., Quigg,~C., Reno, M.~H., Sarcevic,~I.,
{\it Astroparticle Physics} {\bf 5}, 81-110, (1996).

\item Glashow,~S.~L.,
{\it Physical Review} {\bf B118}, 316, (1960).

\item Lipari,~P.,
{\it Astroparticle Physics} {\bf 1}, 195, (1993).

\item Protheroe,~R.~J.,
in {\it Proceedings of IUA Colloquium 163,
Accretion Phenomena and Related Outflows},
edited by D. Wickramasinghe et al., in press, (1996). \\
astro-ph/9607165

\item Rhode,~W. {et al.},
{\it Astroparticle Physics} {\bf 4}, 217, (1996).

\item Stecker,~F.~W., Salamon,~M.~H.,
{\it Space Science Review}, in press, (1995). \\
astro-ph/9501064
 
\item Szabo,~A.~P and Protheroe,~R.~J,
{\it Astroparticle Physics} {\bf 2}, 375, (1994).

\end{list}
\end{document}